\documentclass[onecolumn,preprint,showpacs,amsmath,amssymb]{revtex4-1}
\pdfoutput=1
\usepackage{booktabs,graphicx,pstricks,verbatim,url}
\usepackage{graphicx}
\usepackage{dcolumn}
\usepackage{bm}
\usepackage{slashed}

\def\eeq{\end{equation}}
\def\beq{\begin{equation}}
\def\eeq{\end{equation}}
\def\ba{\begin{eqnarray}}
\def\aa{\end{eqnarray}}

\newcommand{\be}{\begin{eqnarray}}
\newcommand{\ee}{\end{eqnarray}}
\newcommand{\lsim}{\stackrel{<}{\sim}}
\newcommand{\gsim}{\stackrel{>}{\sim}}

\def\ltap{\ \raise.3ex\hbox{$<$\kern-.75em\lower1ex\hbox{$\sim$}}\ }
\def\gtap{\ \raise.3ex\hbox{$>$\kern-.75em\lower1ex\hbox{$\sim$}}\ }
\def\lsim{\ \raise.3ex\hbox{$<$\kern-.75em\lower1ex\hbox{$\sim$}}\ }
\def\gsim{\ \raise.3ex\hbox{$>$\kern-.75em\lower1ex\hbox{$\sim$}}\ }
\def\bea{\begin{eqnarray}}
\def\eea{\end{eqnarray}}

\begin{document}
\input{epsf}

\title{Searching for dilepton resonances below the $Z$ mass at the LHC}

\author{Isaac Hoenig} 
\author{Gabriel Samach}
\author{David Tucker-Smith}
\affiliation{Department of Physics, Williams College, Williamstown, MA 01267}

\date{\today}
\begin{abstract}

We consider LHC searches for dilepton resonances in an intermediate mass range, $\sim 10 -80$ GeV.  We adopt a kinetically mixed $Z'$ as an example of weakly coupled new physics that might have evaded detection at previous experiments  but  which could still be probed  by LHC dilepton spectrum measurements in this mass range.  Based on Monte Carlo simulations, we estimate that existing data from the 7 and 8 TeV LHC could be used to test values of the kinetic mixing parameter $\epsilon$ several times smaller than precision electroweak upper bounds, were an appropriate analysis to be carried out by one of  the experimental collaborations.  

\end{abstract}

\maketitle

\section{Introduction}

The appearance of a new dilepton resonance at the Large Hadron Collider (LHC) would mean the discovery of physics beyond the Standard Model (SM).  Both ATLAS and CMS have searched for dilepton resonances, focusing mainly on the high-mass region above $\sim 100$ GeV.  Here we argue that there is something to be gained by searching  at lower masses -- we focus on the  $\sim 10 -80$ GeV mass range, above the masses of the heaviest hadronic bound states and below the $Z$ peak.  To explore whether LHC searches in this mass range could probe parameter space not already excluded by previous experiments we adopt a particular model for a  weakly coupled dilepton resonance:  a kinetically mixed $Z'$ \cite{Holdom:1985ag}.  A kinetically mixed $Z'$ is a simple addition to the SM.  It is also a central feature in certain models of dark matter 
\cite{Boehm:2003hm,
Pospelov:2007mp,
Foot:2008nw,
ArkaniHamed:2008qn,
An:2009vq,
Chun:2010ve,
Andreas:2011in,
Davoudiasl:2013jma,
Cline:2014dwa}.

For our purposes, the kinetically mixed $Z'$ scenario is parameterized by the $Z'$ mass $M_{Z'}$ and the kinetic mixing parameter $\epsilon$, which controls  the coupling strength of the $Z'$ to SM particles.   
We use Monte Carlo simulations to estimate the potential sensitivity of specialized analyses of existing 7 and 8 TeV data  to a kinetically mixed $Z'$, and to estimate the potential sensitivity of 14 TeV data in the longer term.  
We also investigate whether the CMS analysis of the dilepton invariant-mass spectrum at 7 TeV~\cite{Chatrchyan:2013tia}
implies interesting upper bounds on $\epsilon$ for various intermediate $M_{Z'}$ values, although this exercise is no substitute for an analysis carried out by an experimental collaboration. 

Our main results are summarized in Figure~\ref{fig:combined_results}.  We find that CMS results presented in Ref.~\cite{Chatrchyan:2013tia} imply upper bounds on $\epsilon$ that  lie below  precision electroweak upper bounds in the $30-70$ GeV mass range.  More specialized analyses of existing data  could probe $\epsilon$ values as small as a factor of $\sim 5$ below precision electroweak upper bounds.  The analyses we test start with standard muon $p_T$ cuts used in CMS analyses (14 and 9 GeV for the dimuon pair for $\sqrt{s} = 7$ TeV, and 20 and 10 GeV for $\sqrt{s} = 8$ TeV), but would still be sensitive to $\epsilon \lsim 10^{-2}$ for $Z'$ masses as small as  $M_{Z'} \sim 12$ GeV, although for $M_{Z'}$ below 14 GeV the CMS search for a light pseudoscalar decaying to $\mu^+ \mu^-$ already implies exclusion down to significantly smaller values of $\epsilon$ \cite{Chatrchyan:2012am}.  

We focus on a kinetically mixed $Z'$, but can conclude more generally that, even for dilepton resonances below $M_Z$, the LHC can compete with previous experimental probes. Our results suggest that it would be worthwhile for  LHC experimental collaborations to carry out analyses searching for dilepton resonances in the intermediate mass range considered here.    One can interpret  our sensitivity estimates in the context of a different model by using the cross sections of Table~\ref{tab:cross_sections} and the fact that the production cross section is proportional to $\epsilon^2$, to convert the $\epsilon$ values of Figure~\ref{fig:combined_results} to $\sigma(pp \rightarrow Z' \rightarrow l^+ l^-)$ values.  

In the following section we review the kinetically mixed $Z'$ model and current experimental constraints.  In Section~\ref{sec:7TeV} we describe our simulation and analysis methods and present sensitivity estimates for 7 TeV data.   In Section~\ref{sec:8TeV} we present sensitivity estimates for 8 TeV data and, under certain assumptions,
 for the 14 TeV LHC.

\section{A kinetically mixed $Z'$}
\label{sec:model}

The model we consider is defined by the Lagrangian
\be
{\mathcal L} ={\mathcal L}_{SM}-\frac{1}{4} B'_{\mu \nu} B'^{\mu \nu} - \frac{\epsilon}{2} B_{\mu \nu} B'^{\mu \nu} + \frac{M_{B'}^2}{2} B'_{\mu } B'^{\mu } + \cdots, 
\ee
where ${\mathcal L}_{SM}$ is the SM Lagrangian, $B_{\mu \nu}$ is the hypercharge field strength tensor, and $B'_{\mu \nu}$ is the field strength tensor for a new  $U(1)'$ gauge field.  The mass for the $U(1)'$ gauge field, $M_{B'}$, can be generated by the vacuum expectation value of a SM-singlet scalar field $\Phi$ charged under the $U(1)'$, whose Lagrangian terms we do not write explicitly.  The phenomenology of $\Phi$ can be interesting in its own right, but we do not consider it here. The collider implications of $U(1)'$ gauge bosons in this type of setup have previously been considered in Refs.~\cite{
Babu:1997st,
Rizzo:1998ut,
Strassler:2006im,
Kumar:2006gm,
Chang:2006fp,
Feldman:2007wj,
ArkaniHamed:2008qp,
Cassel:2009pu,
Batell:2009yf,
Essig:2009nc,
Bjorken:2009mm,
Batell:2009di,
Hook:2010tw,
Williams:2011qb,
Frandsen:2012rk,
Toro:2012sv,
Barger:2012ey,
Jaeckel:2012yz,
Davoudiasl:2013aya,
Cline:2014dwa}, 
for example.    

The  neutral $SU(2)$, $U(1)$ and $U(1)'$ gauge fields $W^3_\mu$, $B_\mu$, and $B'_\mu$ are related to mass-eigenstate gauge fields with diagonal and canonically normalized kinetic terms, $A_\mu$, $Z_\mu$, and $Z'_\mu$, by
\be
   \begin{pmatrix} 
      B_\mu  \\
      W^3_\mu \\
      B'_\mu
   \end{pmatrix}
   =
    \begin{pmatrix} 
      c_w & -\left( s_w c_z + \frac{\epsilon}{\sqrt{1-\epsilon^2}} s_z\right)  & \left( s_w s_z - \frac{\epsilon}{\sqrt{1-\epsilon^2}} c_z\right) \\
  s_w & c_w c_z & -c_w s_z \\
 0  & \frac{1}{\sqrt{1-\epsilon^2}}s_z & \frac{1}{\sqrt{1-\epsilon^2}} c_z
   \end{pmatrix}  
      \begin{pmatrix} 
      A_\mu  \\
      Z_\mu \\
      Z'_\mu
   \end{pmatrix},
   \label{eqn:transformation}
\ee
where $c_w$ and $s_w$ are cosine and sine of the weak mixing angle, and $c_z$ and $s_z$ are cosine and sine of the angle that parameterizes $Z-Z'$ mixing, determined by
\be
\tan 2 \theta_z = \frac{2 \epsilon \sqrt{1-\epsilon^2} \;s_w {\overline M}_Z^2}{{\overline M}_Z^2(1-\epsilon^2 -s_w^2 \epsilon^2)-M_{B'}^2}.
\ee
Here ${\overline M}_Z$ is the SM value for the mass of the $Z$ boson, {\it i.e.} its value in the $\epsilon \rightarrow 0$ limit. 

One can use the field transformation of Equation~(\ref{eqn:transformation}) and the charges of SM fermions under the SM gauge group to calculate the couplings of SM fermions to the $Z'$. 
If both 
\be
\epsilon \ll 1 
\quad \quad \quad \quad 
\text{and} 
 \quad \quad \quad \quad
\left|  \frac{ \epsilon s_w {\overline M}_Z^2}{{\overline M}_Z^2-M_{B'}^2} \right| \ll 1
 \label{eqn:smallmix}
\ee
are satisfied, we can work to first order in $\epsilon$, giving
\be
\theta_z \simeq  \frac{ \epsilon s_w {\overline M}_Z^2}{{\overline M}_Z^2-M_{B'}^2}.
\ee
In this regime, the couplings of the $Z'$ to SM fermions are
\be
g_{\bar{f}f Z'} \simeq -\epsilon \left( \frac{ {\overline M}_Z^2 c_w e Q_f -M_{B'}^2 g_y Y_f}{{\overline M}_Z^2-M_{B'}^2} \right),
\label{eqn:approxcouplings1}
\ee
where $e$ and $g_y$ are the electromagnetic and hypercharge gauge couplings, $Q_f$ and $Y_f$ are the electric charge and hypercharge of the fermion, and 
where we use the notation
\be
{\mathcal L} \supset \sum_f g_{\bar{f}f Z'} \overline{f} \gamma^\mu f Z'_\mu,
\ee 
with $f = e_L, \; e_R,$ and so on.    
When either $M_{B'} \ll {\overline M}_Z$ or $M_{B'} \gg {\overline M}_Z$ is satisfied we can further approximate Equation~(\ref{eqn:approxcouplings1}) as
\be
g_{\bar{f}f Z'} \simeq
\begin{cases}
-\epsilon c_w e Q_f  &  M_{B'} \ll {\overline M}_Z\\
-\epsilon g_y Y_f & M_{B'} \gg {\overline M}_Z.
\end{cases}
\ee


 In the context of a dark matter model,  the $Z'$ might have a sizable branching ratios to dark-sector final states.  In this work  we assume  that the $Z'$ decays exclusively to SM states.  This possibility fits in naturally with viable dark matter models, even for small values of $\epsilon$.  For example, if the dark matter is a fermion $\chi$ charged under $U(1)'$,  $Z'$ decays to $\chi$ pairs are forbidden for $M_\chi> M_{Z'}/2$, and for $M_\chi> M_{Z'}$ the $\chi$ relic abundance  can be regulated by the annihilation process $\chi \overset{\tiny{(-)}}{\chi} \rightarrow Z' Z'$ for an appropriate value of the $U(1)'$ gauge coupling $g'$ \cite{Hooper:2012cw}.    For the parameter region relevant to the collider studies considered in this paper ($M_Z'$ in the $\sim 10-80$ GeV range and $\epsilon > 10^{-3}$) 
results from LUX \cite{Akerib:2013tjd} rule out this $\chi \overset{\tiny{(-)}}{\chi} \rightarrow Z' Z'$ scenario for the case where the dark matter is a Dirac fermion, but the Majorana case is viable.  A Majorana mass can be generated by $\Phi-\chi$ interactions for suitably chosen $U(1)'$ charges.  

Although dark matter offers one motivation to search for weakly coupled dilepton resonances, our results do not depend on the possible connection to  dark matter.  The key assumption is that the width of the dilepton resonance is dominated by SM final states.  

Experimental and observational constraints on the kinetically mixed $Z'$ scenario are summarized in Ref.~\cite{Jaeckel:2013ija}.  In  the  $\sim 10-80$ GeV mass range that interests us, precision electroweak constraints require $\epsilon \lsim (2-3)\times 10^{-2}$ \cite{Hook:2010tw}.  
If the $Z'$ decays dominantly to SM states, as assumed here, there are also constraints from PEP, PETRA, and TRISTAN  measurements of the $e^+ e^- \rightarrow \text{hadrons}$ cross sections at various center-of-mass energies.  As shown in Ref.~\cite{Hook:2010tw}, these constraints are stronger than those from precision electroweak measurements at particular values of $M_{Z'}$ close to the experiments' center-of-mass energies, but they do not cover the mass range continuously.  

\begin{figure}[htbp]
\begin{center}
\includegraphics[width=4in]{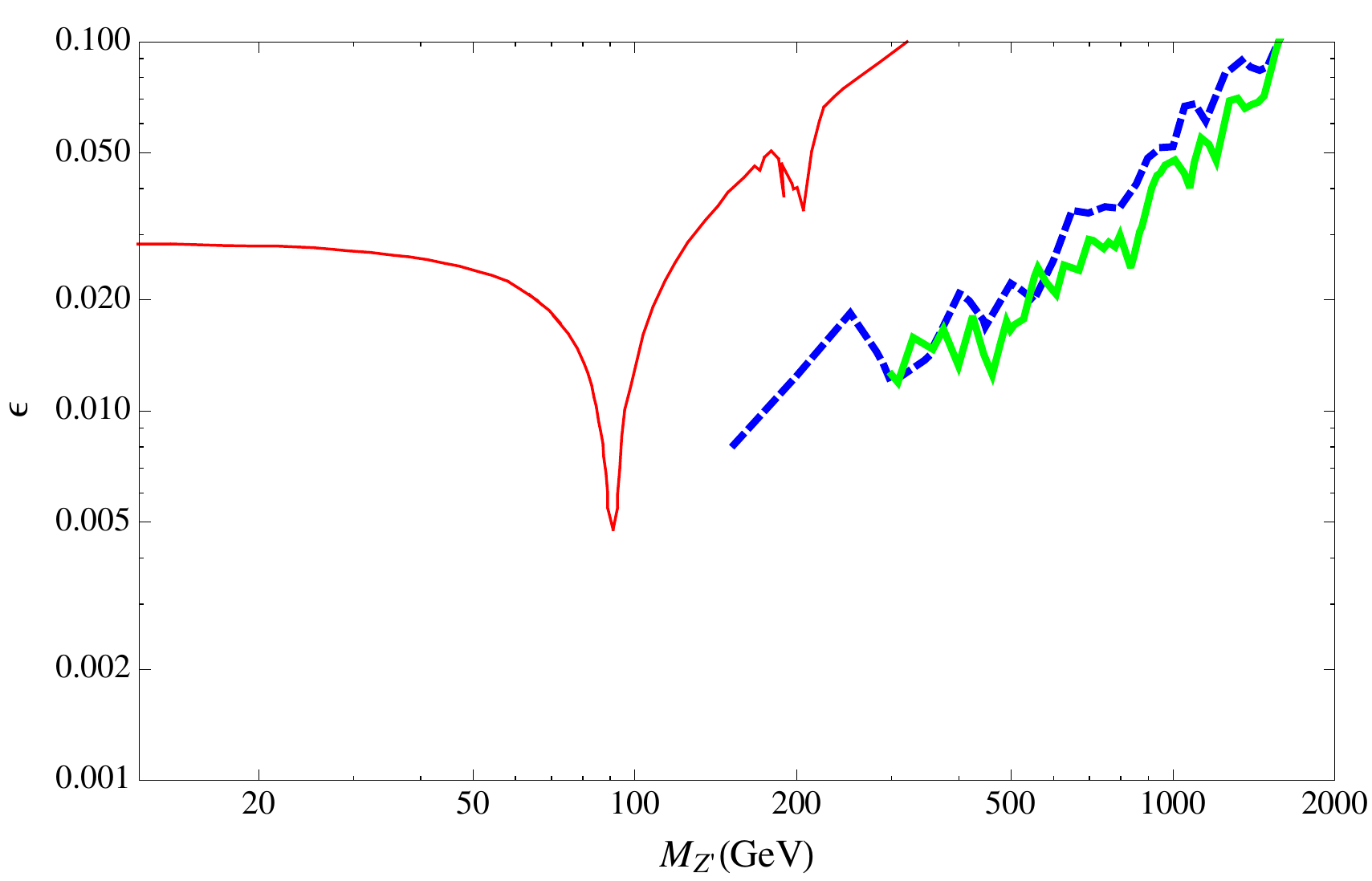}
\caption{95\% CL upper bounds on $\epsilon$ from ATLAS \cite{ATLAS8TeVdilepton} (dashed blue line) and CMS \cite{CMS8TeVdilepton} (thick green line) searches for high-mass dilepton resonances, along with the precision electroweak constraint taken from Ref.~\cite{Hook:2010tw} (thin red line).}
\label{fig:existing_constraints}
\end{center}
\end{figure}

At higher masses, LHC searches for high-mass dilepton resonances can be used to constrain the kinetically mixed $Z'$ scenario.  These constraints were derived in Ref.~\cite{Jaeckel:2012yz} using 7 TeV LHC results.  
Figure \ref{fig:existing_constraints} shows the upper bounds on $\epsilon$ implied by ATLAS \cite{ATLAS8TeVdilepton} and CMS \cite{CMS8TeVdilepton} 8 TeV results.   
To obtain these constraints we calculate the cross section for $pp \rightarrow Z' \rightarrow l^+ l^-$ at leading order using Madgraph \cite{Alwall:2011uj} with CTEQ66 \cite{Kretzer:2003it,Nadolsky:2008zw} parton distribution functions, and apply a mass-dependent NNLO K-factor determined using the ZWPROD code \cite{Hamberg:1990np,vanNeerven:1991gh}.

Given how successful LHC searches for high-mass dilepton resonances have been at probing new parameter space for the kinetically mixed $Z'$ scenario, a natural question is whether LHC data can be used to probe new parameter space for $M_{Z'} < M_Z$.  The results of a CMS search for a light pseudoscalar Higgs in the dimuon channel, which relied on a specialized trigger,  imply a limit on $\epsilon$ for $M_{Z'}$ in the $5.5 - 14$ GeV mass range \cite{Chatrchyan:2012am} (for an earlier ATLAS analysis, see Ref.~\cite{ATLAS:2011cea}).  For example, assuming equal acceptances for the $Z'$ and the pseudoscalar, we find that the CMS upper limit implies $\epsilon< 2\times 10^{-3}$ for $M_{Z'} =12$ GeV. 
  In the following sections we consider whether LHC studies of the dimuon invariant mass distribution could in principle be used to probe the kinetically mixed $Z'$ model for a broader range of intermediate $Z'$ masses.   From simulations with standard cuts taken from existing analyses,  we find that current data could be used to probe $\epsilon$ values well below the precision electroweak  limit for much of the mass range below $M_Z$, although not down to the $\epsilon \sim 2\times 10^{-3}$ level probed by the CMS pseudoscalar search for masses below 14 GeV.    It is possible that Tevatron data could also be used to probe $\epsilon$ values below precision electroweak upper bounds. Our rough estimates suggest that it is unlikely that  the sensitivity of Tevatron data would exceed that of LHC data, but we have not investigated the Tevatron sensitivity in detail.

\section{Potential sensitivity of 7 TeV LHC data} \label{sec:7TeV}
\subsection{simulation methods}

We perform Monte Carlo simulations of dimuon production to estimate the sensitivity of LHC data to the kinetically mixed $Z'$ model for $M_{Z'} < M_Z$.  We implement the model in FeynRules \cite{Christensen:2008py} and simulate both the $Z'$ signal and the dominant Drell-Yan background using MadGraph5 \cite{Alwall:2011uj} and Pythia 6.4 \cite{Sjostrand:2006za}, with Madgraph's implementation of MLM matching \cite{Alwall:2007fs} turned on, and up to two jets included at matrix element level.  We approximate detector resolution effects by performing a Gaussian smearing of the $p_T$'s of the muons using the smearing function
\be
\frac{\sigma_{p_T}}{p_T} = \begin{cases}
0.03 & |\eta|<1.5 \text{ and }  p_T<200\\
0.04 & |\eta| \ge 1.5 \text{ and }  p_T < 200\\
0.05 &  p_T \ge 200\\
\end{cases}
\ee
This smearing function is consistent with the settings in the default CMS card for the Delphes 3.0 fast detector simulator, which have been shown to give reasonable agreement with data  \cite{deFavereau:2013fsa}.  
The sizes of our background Monte Carlo samples correspond to effective luminosities roughly a factor of ten times larger than the luminosities used for the associated analyses. 

\begin{figure}[h]
\begin{center}
\includegraphics[width=3in]{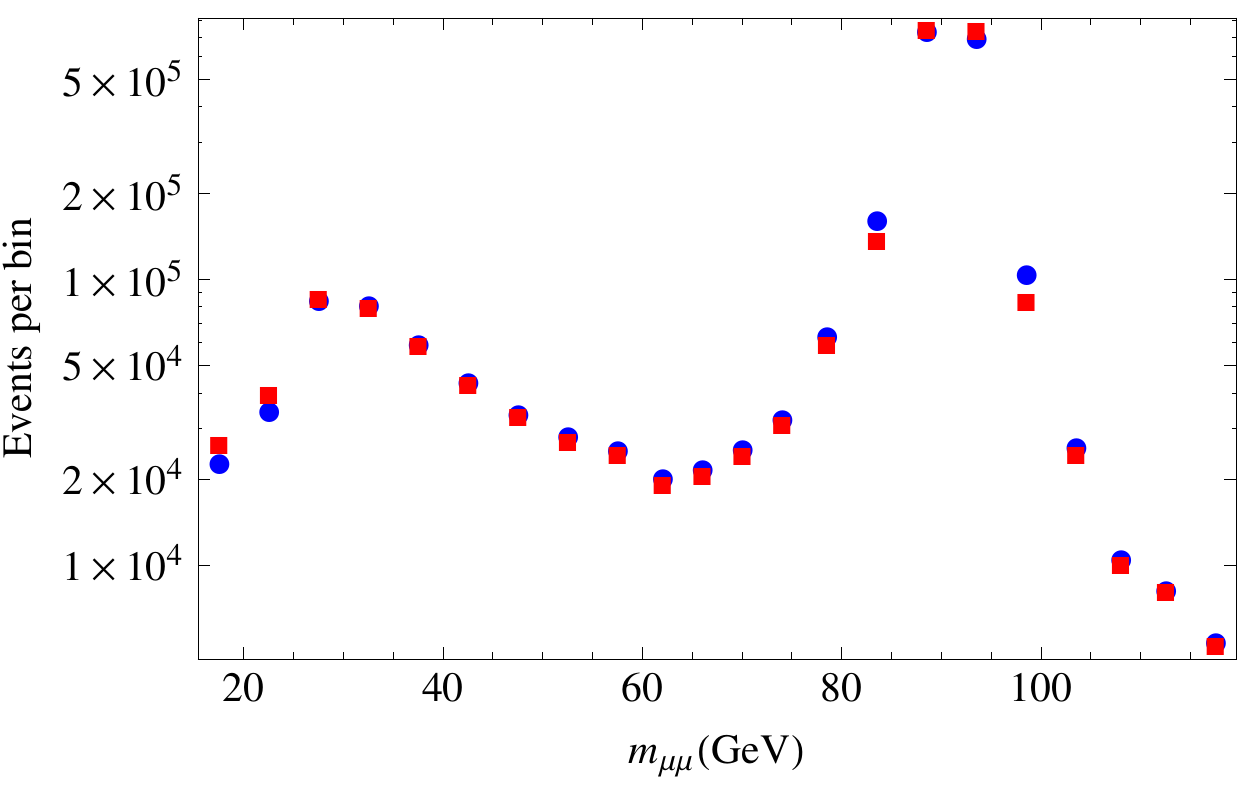}
\hspace{0.2in}
\includegraphics[width=3in]{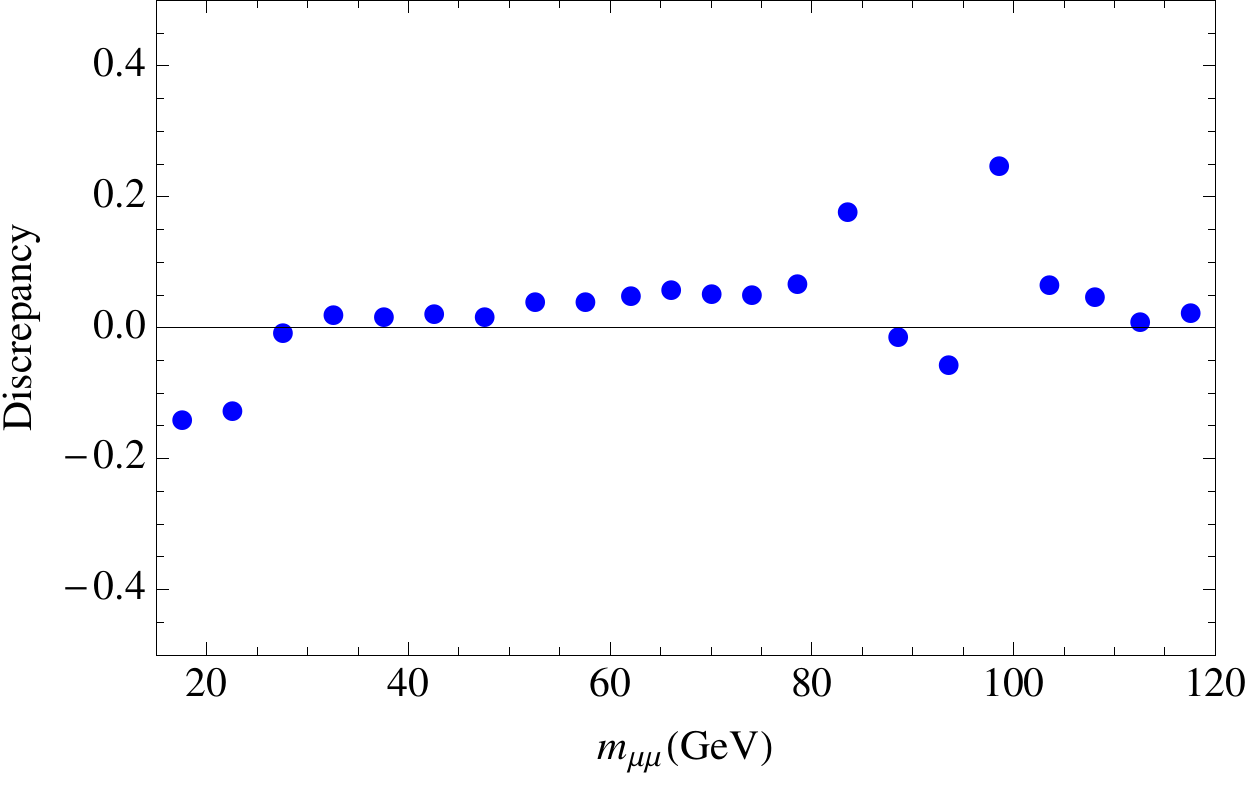}
\caption{Left: with $\sqrt{s} = 7$ TeV and an integrated luminosity of 4.5 fb$^{-1}$, a comparison between the Drell-Yan dimuon invariant-mass distributions predicted by CMS simulations (red squares) and by our simulations (blue circles).  The muons are required to have $\left(p_T \right)_1 > 14$ GeV, $\left(p_T \right)_2>9$ GeV, and $|\eta|<2.4$. We rescale our $m_{\mu \mu}$ distribution to give the same total counts as the CMS simulation in the $60-120$ GeV mass range.  Right:  discrepancies between our simulated bin counts and those of CMS.}
\label{fig:simulation_compare}
\end{center}
\end{figure}

Drell-Yan production in the mass range that interests us has been studied at  7 TeV by ATLAS~\cite{Aad:2014qja} and CMS \cite{Chatrchyan:2013tia}.
In this section we will frequently refer to the CMS analysis, which uses an integrated luminosity of 4.5 fb$^{-1}$. 
Figure 1 of that paper shows that, for the selection cuts used,  Drell Yan production accounts for approximately 90\% of dimuon production in the mass range that interests us.  The CMS analysis requires the leading and subleading muons to have $\left(p_T\right)_1 > 14$ GeV and $\left(p_T\right)_2>9$ GeV, respectively, with both muons' pseudorapidities satisfying $|\eta|<2.4$.  For these cuts,  Figure~\ref{fig:simulation_compare} compares the dimuon invariant-mass distribution produced by our simulations with that produced by the CMS full simulation of the Drell-Yan process, as reported in Ref.~\cite{Chatrchyan:2013tia}.  
We rescale our distribution to produce the same number of events as the CMS simulation in the $60-120$ GeV mass range.  We find agreement at the 5\% level for the bins in the $25-76$ GeV mass range.   Larger discrepancies approaching 15\% are evident in the two lowest-mass bins, where higher order effects are most important.  Larger discrepancies are also evident around the $Z$ mass, where the bin counts depend sensitively on the muon momentum resolution.  Our simulations give a $Z$ peak that is slightly too short and broad, suggesting that we are not unreasonably optimistic in our estimation of the resolution.


A light kinetically mixed $Z'$ can also be searched for in the dielectron channel, and combining dimuon and dielectron results would likely lead to improved sensitivity.  For simplicity we restrict our attention to the dimuon channel, in part because, as shown in Figure 3 of Ref.~\cite{Chatrchyan:2013tia}, the efficiency in the dielectric channel is significantly lower in the invariant mass range that interests us.  


\subsection{A benchmark scenario: $M_{Z'} = 50$ GeV, $\epsilon = 0.02$  }

To establish that a $Z'$ signal would be detectable at the LHC for parameters consistent with precision electroweak bounds,  we adopt the parameter point ($M_{Z'} = 50$ GeV, $\epsilon = 0.02$), which lies slightly below the precision electroweak exclusion contour in Figure \ref{fig:existing_constraints}.  We normalize our background invariant-mass distribution to data, using the number of events reported in the $60-120$ GeV mass range in the CMS analysis of Ref.~\cite{Chatrchyan:2013tia}.  
To normalize the signal
we use the matched cross section calculated by MadGraph with CTEQ66 parton distributions \cite{Kretzer:2003it,Nadolsky:2008zw} and, in addition to selection cuts, we apply an efficiency of 75\%, consistent with or slightly below the efficiencies reported in Ref.~\cite{Chatrchyan:2013tia}.  
Signal cross sections calculated by MadGraph for various $Z'$ masses are given in Table~\ref{tab:cross_sections}.

\begin{figure}[htbp]
\begin{center}
\includegraphics[width=4in]{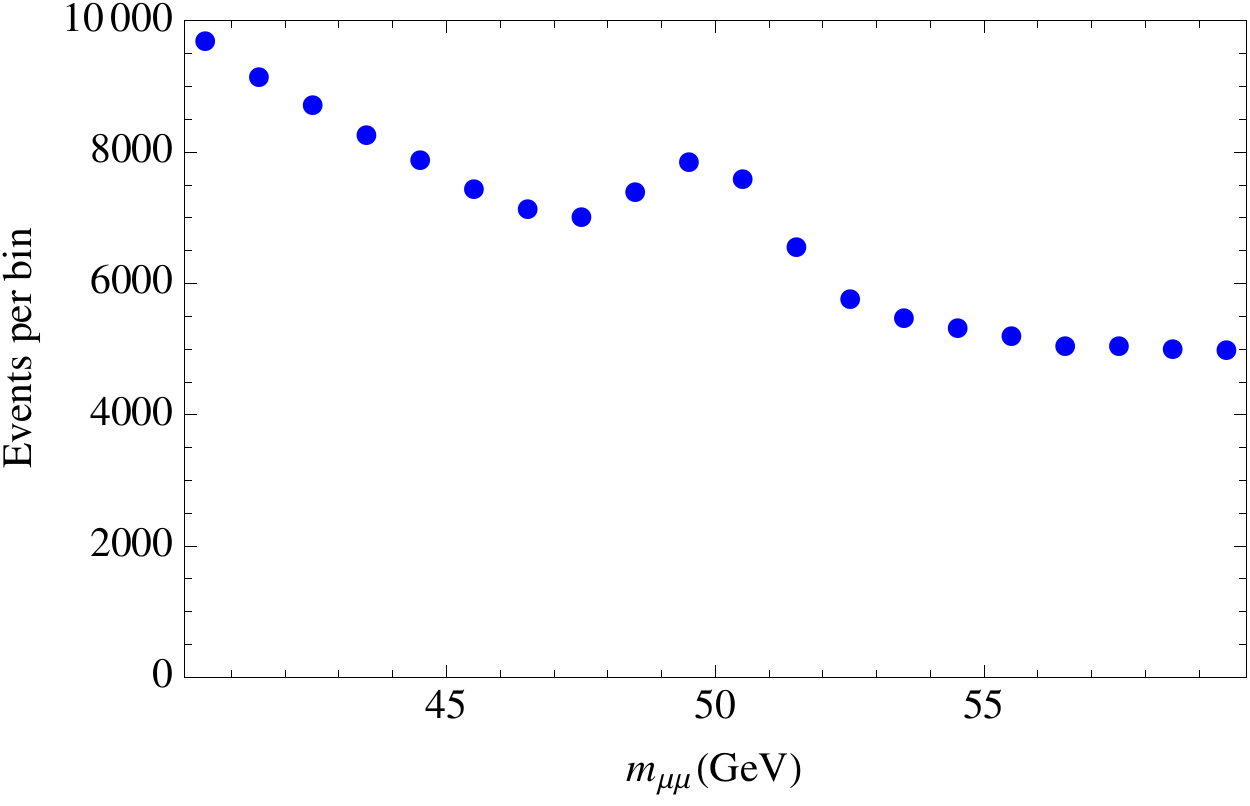}
\caption{With $\sqrt{s} = 7$ TeV and an integrated luminosity of 4.5 fb$^{-1}$, simulated signal plus background for the benchmark parameter point $M_{Z'} = 50$ GeV and $\epsilon = 0.02$.  The muons are required to have $\left(p_T \right)_1 > 14$ GeV, $\left(p_T \right)_2>9$ GeV, and $|\eta|<2.4$. We normalize background to data and apply a 75\% efficiency to signal.}
\label{fig:benchmark_signal}
\end{center}
\end{figure}

The invariant mass distribution for signal plus background for $M_{Z'} = 50$ GeV and $\epsilon = 0.02$, using 1 GeV bins, is shown in Figure \ref{fig:benchmark_signal}.
For the parameters of this benchmark scenario, the $Z'$ signal would be clearly visible in 7 TeV data, motivating us to investigate further the potential sensitivity of the LHC to the kinetically mixed $Z'$ model in the $M_{Z'} < M_Z$ regime.

\subsection{Models for the dimuon invariant mass distributions}
\label{sec:background_model}
To estimate the $\epsilon$ reach of 7 TeV LHC data, we adopt models for the background and signal invariant mass distributions.  We model the signal using the invariant mass bin counts produced by our Monte Carlo simulations.  Throughout we apply a $75\%$ efficiency to the signal.  
For the background we employ a more flexible model that allows for deviations of actual SM $m_{\mu \mu}$ distributions from those predicted by simulations. 
  The bin counts predicted by our background model  are
\begin{equation}
b_i(\theta,\delta) = p_i(\theta)( 1 + \delta_i),
\end{equation}
where $p_i$ represents a fifth-order polynomial with coefficients $\theta$, and the $\delta_i$ are additional nuisance parameters, one for each bin.  When we incorporate a possible signal 
 the predicted bin counts become
\be
\nu_i (\mu, \theta,\delta) = \mu s_i+b_i(\theta, \delta),
\ee
where $s_i$ are the Monte-Carlo derived signal bin counts for particular pair of ($\epsilon$, $M_{Z'} $) values, and $\mu$ controls the signal strength.

We quantify the compatibility of signal strength $\mu$ with data $n_i$ by
\be
\chi^2(\mu) =\min_{\{\theta,\; \delta\}} 
\left\{
\sum_i \left[      \frac{\left[ n_i - \nu_i (\mu, \theta,\delta) \right]^2}{\nu_i (\mu, \theta,\delta)}     +\left( \frac{\delta_i}{\kappa} \right)^2 \right]
\right\}.
\label{eqn:chisquared}
\ee
In this expression, the background parameters $\theta$ and $\delta$ are set to their best-fit values for signal strength $\mu$.  That is, they are chosen to minimize the quantity in brackets.  The $\delta$ parameters are effectively associated with fake measurements of uncertainty $\kappa$.  In the $\kappa \rightarrow 0$ limit the background model reduces to a fifth-order polynomial.  We describe how we choose $\kappa$  values below.  

\begin{figure}[h]
\begin{center}
\includegraphics[width=4in]{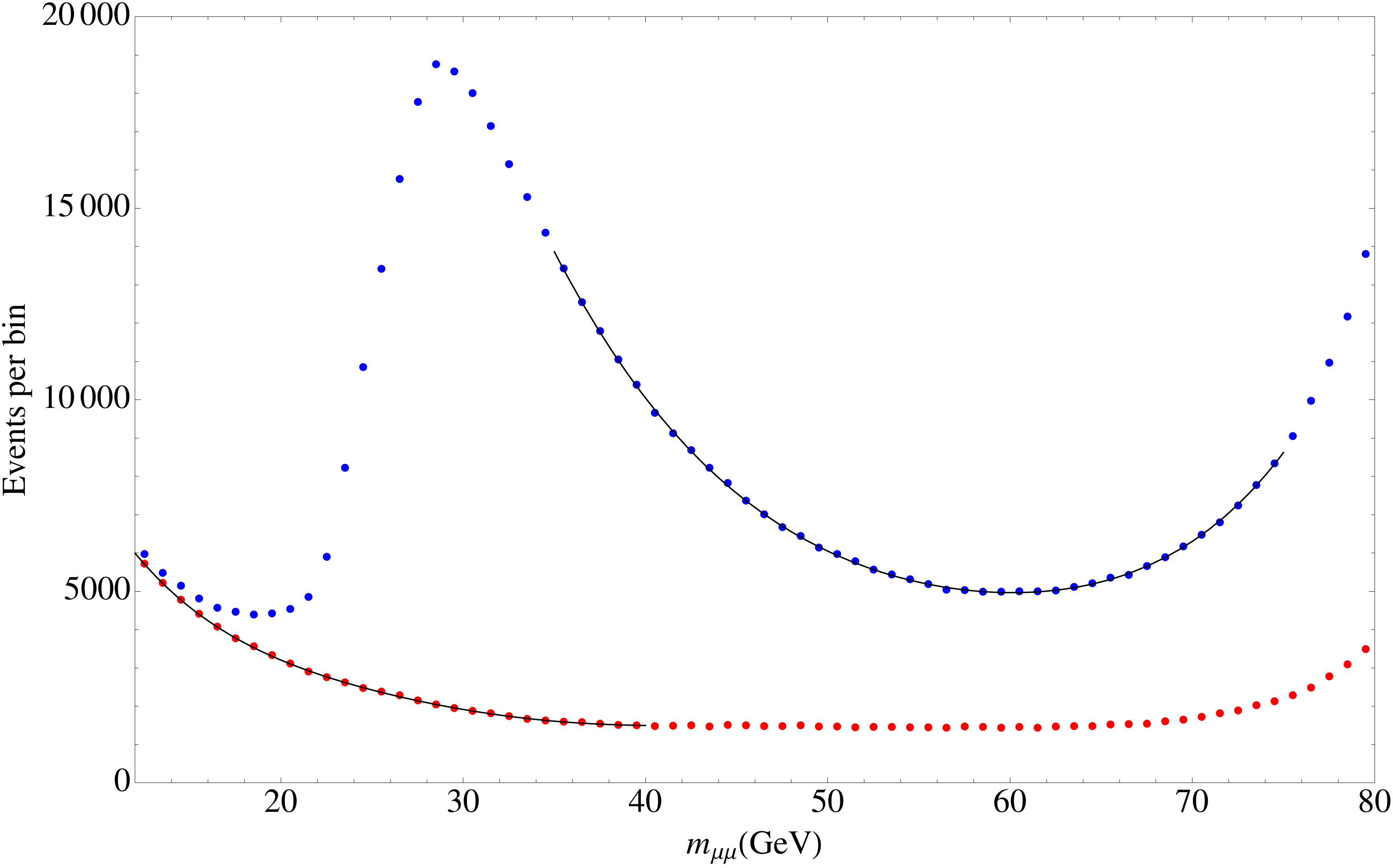}
\put(-170,130){\tiny no $\left( p_T \right)_{\mu \mu}$ cut}
\put(-220,42){\tiny $\left( p_T \right)_{\mu \mu}>23$ GeV}
\caption{For $\sqrt{s} = 7$ TeV and an integrated luminosity of 4.5 fb$^{-1}$, our simulated Drell-Yan $m_{\mu\mu}$ distributions with and without the $\left( p_T \right)_{\mu \mu}$ cut, along with best-fit fifth-order polynomial curves. The muons are required to have $\left(p_T \right)_1 > 14$ GeV, $\left(p_T \right)_2>9$ GeV, and $|\eta|<2.4$.}
\label{fig:polynomial_fit_plot}
\end{center}
\end{figure}

Taking the signal strength to zero and neglecting the $\delta$ parameters (or equivalently, sending $\kappa \rightarrow 0$), we can identify the $\theta$ parameters that give the best fifth-order polynomial fit to our Drell-Yan Monte Carlo $m_{\mu \mu}$ distribution.  
Figure~\ref{fig:polynomial_fit_plot} shows this fit for the $35-75$ GeV mass range.  
At lower masses, matters are complicated by the fact that the muon $p_T$ cuts produce a peak in the invariant mass distribution near $m_{\mu \mu} =  30$ GeV. Dimuon masses below $\sim 30$ GeV require the dimuon system to recoil off of a jet or jets in order to satisfy $\left(p_T\right)_1>14$ GeV and $\left(p_T\right)_2>9$ GeV . 

A cut on $(p_T)_{\mu \mu}$, the transverse momentum of the dimuon system, can be used to probe $Z'$ masses below $\sim 30$ GeV.  We choose the cut on  $(p_T)_{\mu \mu}$ to be equal to the sum of the two individual muon $p_T$ thresholds, 23 GeV for the muon $p_T$ cuts used in the CMS analysis of Ref.~\cite{Chatrchyan:2013tia}.  With this choice, an event with very small $m_{\mu \mu}$ that passes the individual muon $p_T$ cuts will also pass the cut on  $(p_T)_{\mu \mu}$.  Here and for all of our analyses we require that the separation between the $\mu^+$ and $\mu^-$ satisfies $\Delta R > 0.3$.  
Figure~\ref{fig:polynomial_fit_plot} shows the best fifth-order polynomial fit for the $12-40$ GeV mass range when this additional cut is applied to to our Drell-Yan Monte Carlo sample.

To calculate a $p$-value for the background-model fit to a given data set $n_i$, we first use Equation~(\ref{eqn:chisquared})
to calculate $\chi^2(0)$ and extract best-fit background parameters $\theta_{(0)}$ and $\delta_{(0)}$.
We define the $p$-value for the fit to be
\be
p = \int_{\chi^2(0)}^\infty \!  f(x; n_d) \; dx,
\ee
where $ f(x; n_d)$ is the probability distribution function for the chi squared distribution with $n_d$ degrees of freedom, and we take $n_d$ equal to the number of bins minus six, the number of parameters for the fifth-order polynomial \footnote{Starting with pseudo data sets generated by statistically  fluctuating the background-model prediction $b_i(\theta_{(0)},\delta_{(0)})$, and associated sets of  $\delta^\text{fake}_i$ generated by fluctuating about ${\delta_{(0)}}_i$ with standard deviation $\kappa$, the $p$-value that we have defined is  the fraction of those pseudo data/ $\delta^\text{fake}_i$ sets for which $\chi^2(0)$ is greater than for the original data set.
When calculating $\chi^2(0)$ for the pseudo data sets, one replaces  $\delta_i$ with $\delta_i - \delta^\text{fake}_i$ in Equation~(\ref{eqn:chisquared}).}.

In Section~\ref{sec:bounds_from_CMS_analysis} we estimate bounds on $\epsilon$ for various $Z'$ masses based on the CMS results of Ref.~\cite{Chatrchyan:2013tia}, and in Section~\ref{sec:sensitivity_estimates_7TeV} we estimate the potential sensitivity of 7 TeV LHC data using two hypothetical analyses with 1-GeV $m_{\mu \mu}$ bins, one with the $\left( p_T \right)_{\mu \mu} > 23$ GeV cut and the other without.  
Tables~\ref{tab:7TeV_CMS_fits}--\ref{tab:7TeV_14_9_23_fits} show the $M_{Z'}$ values tested for each analysis and the associated $m_{\mu \mu}$ ranges.
For each $m_{\mu \mu}$ range, the third column of the table gives the median $p$-value for the fifth-order polynomial fit
to pseudo data generated by statistically fluctuating Monte Carlo expectations for SM  bin counts.  We use the Monte Carlo expectations reported by CMS for the analysis based on the CMS results, and we use our own Monte Carlo simulations for the 1-GeV analyses 

 We choose  $\kappa$ values by imposing two requirements.   First, we require that the median $p$ value for background-model fits to SM Monte-Carlo-generated psuedo data is at least 0.5.  Second, we require that when a signal corresponding to kinetic mixing parameter $\epsilon_{in}$ is injected into these pseudo-data sets, the 95\% confidence level upper bound on $\epsilon$,   $\epsilon_{95}$, is found to be less than $\epsilon_{in}$ for fewer than 5\% of the pseudo-data sets.  Our procedure for calculating $\epsilon_{95}$ for a particular data set is described in the following section.
Because we impose the second requirement independently for each hypothesized value of $M_{Z'}$, different values of $M_{Z'}$ can have different values of $\kappa$.  For each value of $M_{Z'}$, the second requirement is imposed over a  range of $\epsilon_{in}$ values extending down to zero and up to twice the median of the $\epsilon_{95}$ values obtained for the SM-only pseudo-data sets.  The $\kappa$ values obtained by imposing our two requirements fill the fourth column in Tables~\ref{tab:7TeV_CMS_fits}--\ref{tab:7TeV_14_9_23_fits}.

The $\kappa$ parameter is meant to take into account imperfections in the polynomial background model.
 An actual experimental analysis would deal with systematic effects  not considered here, and these effects may make the background $m_{\mu \mu}$ distribution less smooth than our Monte Carlo simulations would suggest.    Along with the $\kappa$ values given in Tables~\ref{tab:7TeV_CMS_fits}--\ref{tab:7TeV_14_9_23_fits}, we present results for $\kappa = 0$, $\kappa=10^{-2}$, $\kappa=2\times 10^{-2}$, and $\kappa=3\times 10^{-2}$ to give a sense of how the potential sensitivity is affected if systematic effects introduce an additional source of bin-to-bin randomness.

\subsection{Statistical procedure}
\label{sec:statistics}
Given a hypothesized value of $M_{Z'}$ and an observed invariant mass distribution $n_i$ -- which could either stand for actual data reported by CMS in Ref.~\cite{Chatrchyan:2013tia}, or for pseudo data generated from our Monte Carlo simulations -- we use the $CL_s$ method \cite{Read:2002hq} to determine $\epsilon_{95}$, the 95\% confidence level upper limit on the kinetic mixing parameter for that value of $M_{Z'}$.  We use a test statistic based on the $\chi^2$ defined in Equation~(\ref{eqn:chisquared}):

\be
q_\mu = 
\begin{cases}
\chi^2(\mu) - \chi^2(\hat{\mu})  
\quad
\quad
\quad\quad\quad
\mu > \hat{\mu}\\
0
\quad\quad\quad\quad\quad\quad\quad\quad\quad\quad \; \;
 \mu \leq \hat{\mu},
\end{cases}
\ee
where $\hat{\mu}$, is the best-fit signal strength.
Using the asymptotic formulae of Ref.~\cite{Cowan:2010js}, we can approximate $CL_s$ as
\begin{equation}
CL_{s} = \frac{1-\Phi(\sqrt{q_{\mu}})}{1-\Phi(\sqrt{q_{\mu}}-\sqrt{q_{\mu,A}})},
\end{equation}
where $\Phi$ is the cummulative distribution of the standard Gaussian and $q_{\mu,A}$ is the value of $q_\mu$ obtained when the data $n_i$ are taken to be equal to $\nu_i(0, \theta_{(0)},\delta_{(0)})$, the expected bin counts for the best-fit background model.  For the purpose of calculating $q_{\mu,A}$,  we replace $\delta_i$ is with $\delta_i - {\delta_{(0)}}_i$ in the $\chi^2$ expression of Equation~(\ref{eqn:chisquared}).
We  define $\epsilon_{95}$ to be the kinetic mixing parameter that gives $CL_s = 0.05$ for $\mu = 1$.

\subsection{Estimates of bounds on $\epsilon$ from the 7 TeV CMS analysis}

\label{sec:bounds_from_CMS_analysis}
Focusing on the $30-76$ GeV bins, the smoothness of the $m_{\mu \mu}$ distribution for the data shown in Figure 1 of the CMS analysis of Ref.~\cite{Chatrchyan:2013tia} implies a constraint on the couplings of a $Z'$ with a mass in this range. We  use the background model and statistical procedure outlined in Sections \ref{sec:background_model} and \ref{sec:statistics} to estimate upper bounds on $\epsilon$ based on the CMS results.  
The $Z'$ masses tested, associated $m_{\mu \mu}$ fit ranges, and $\kappa$ values are shown in Table~\ref{tab:7TeV_CMS_fits}.
\begin{figure}[h]
\begin{center}
\includegraphics[width=4in]{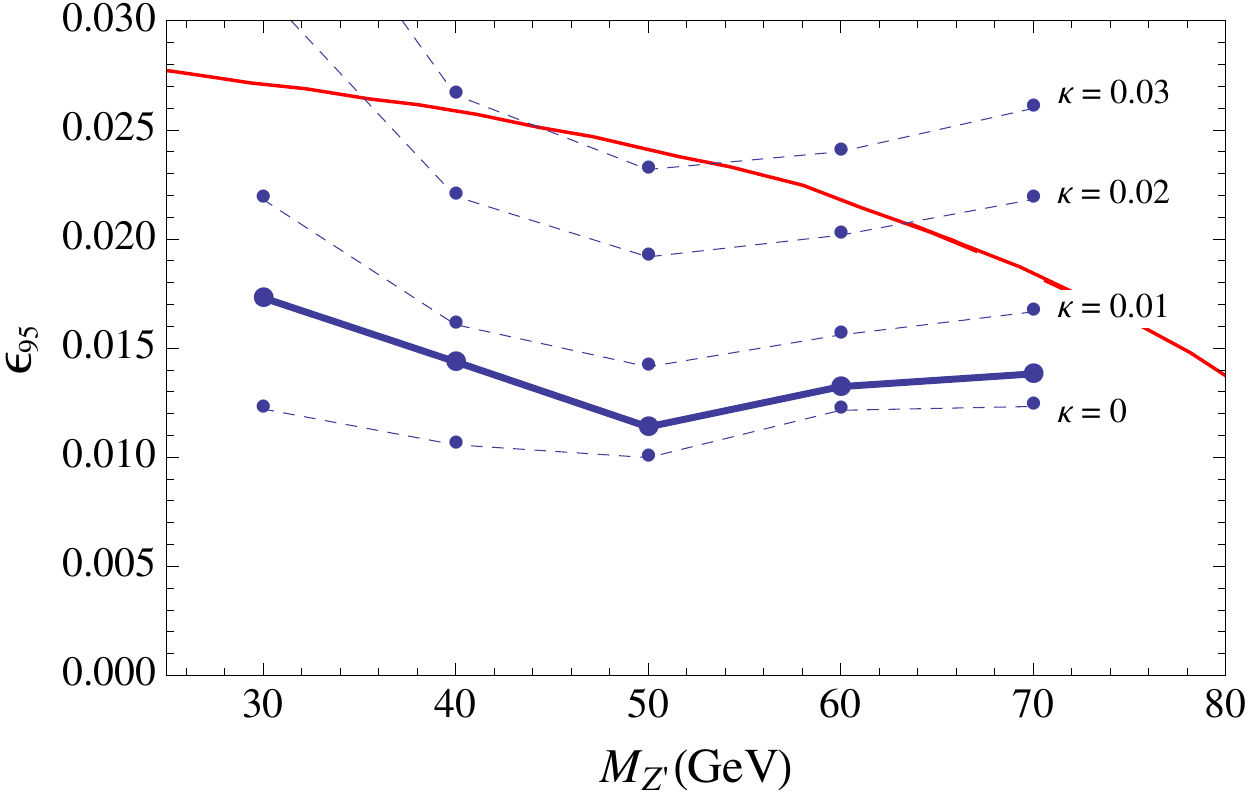}
\caption{Estimated 95\% CL upper bounds on $\epsilon$ based on Ref.~\cite{Chatrchyan:2013tia}, the $\sqrt{s} = 7$ TeV CMS study of the dimuon invariant mass distribution. For the thick blue line, $\kappa$ is set according to the procedure described in Section \ref{sec:background_model}.  The dashed lines show results for other $\kappa$ values.  The thin red line is the precision electroweak constraint taken from Ref.~\cite{Hook:2010tw}.}
\label{fig:results_CMS_analysis}
\end{center}
\end{figure}

Based on the reported data bin counts from Figure 1 of Ref.~\cite{Chatrchyan:2013tia}, we obtain the $\epsilon_{95}$ estimates presented in Figure \ref{fig:results_CMS_analysis}. The estimated $\epsilon$ bounds corresponding to the $\kappa$ values of Table~\ref{tab:7TeV_CMS_fits}  lie below precision electroweak constraints by as much as a factor of two.  Of the five masses tested, the four lowest lie on the boundary between two of the invariant mass bins used in the CMS analysis, meaning that the signal events are roughly equally distributed between two relatively wide mass bins.  This pushes our estimates in the conservative direction, as far as other masses are concerned.

\subsection{Sensitivity estimates for 7 TeV data}

\label{sec:sensitivity_estimates_7TeV}

A more specialized analysis of LHC data could probe lower values of $\epsilon$ and a broader range of $Z'$ masses.   To estimate the potential sensitivity of $\sqrt{s} = 7$ TeV data we calculate median values of $\epsilon_{95}$ for fake data generated from our Drell-Yan Monte Carlo simulations.  Although it would be more interesting for the LHC to find evidence for a light $Z'$ rather than to rule out additional parameter space, we use potential exclusion sensitivity in the absence of new physics as a diagnostic for whether new parameter space can be probed. 

We consider two hypothetical analyses with 1 GeV $m_{\mu \mu}$ bins, both based on the same integrated luminosity as the CMS analysis.   For the first, which
incorporates the same muon $p_T$ and $\eta$ cuts as the CMS analysis,  we use the $Z'$ masses, $m_{\mu \mu}$ fit ranges, and $\kappa$ values from Table~\ref{tab:7TeV_14_9_0_fits}.
The second analysis adds the cut $(p_T)_{\mu \mu}>23$ GeV, and has been tested  with the $Z'$ masses, $m_{\mu \mu}$ fit ranges, and $\kappa$ values from Table~\ref{tab:7TeV_14_9_23_fits}.
We normalize our Monte Carlo simulations so that, in the absence of the $(p_T)_{\mu \mu}$ cut, we get the same number of selected events in  the $60-120$ $m_{\mu \mu}$ range as the CMS analysis, as determined using Figure 1 of Ref.~\cite{Chatrchyan:2013tia}. 

\begin{figure}[h]
\begin{center}
\includegraphics[width=3.in]{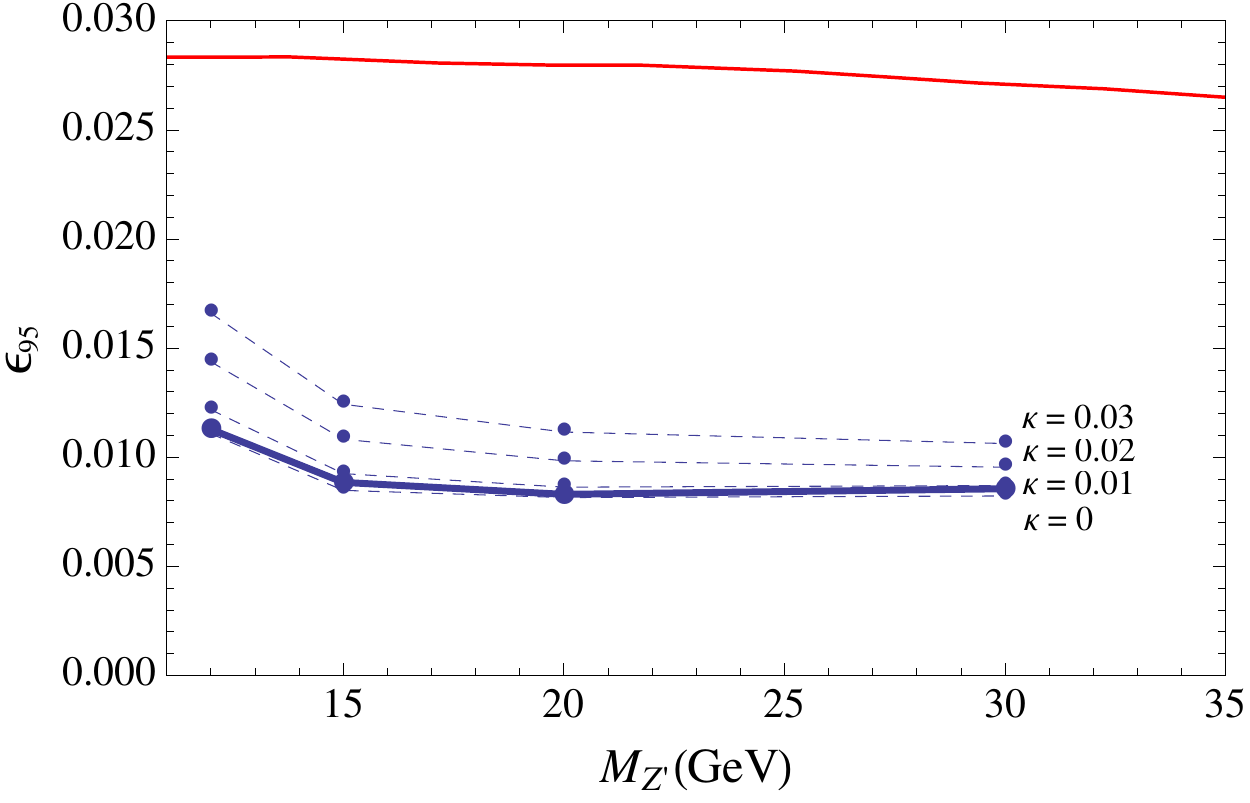}
\includegraphics[width=3.2in]{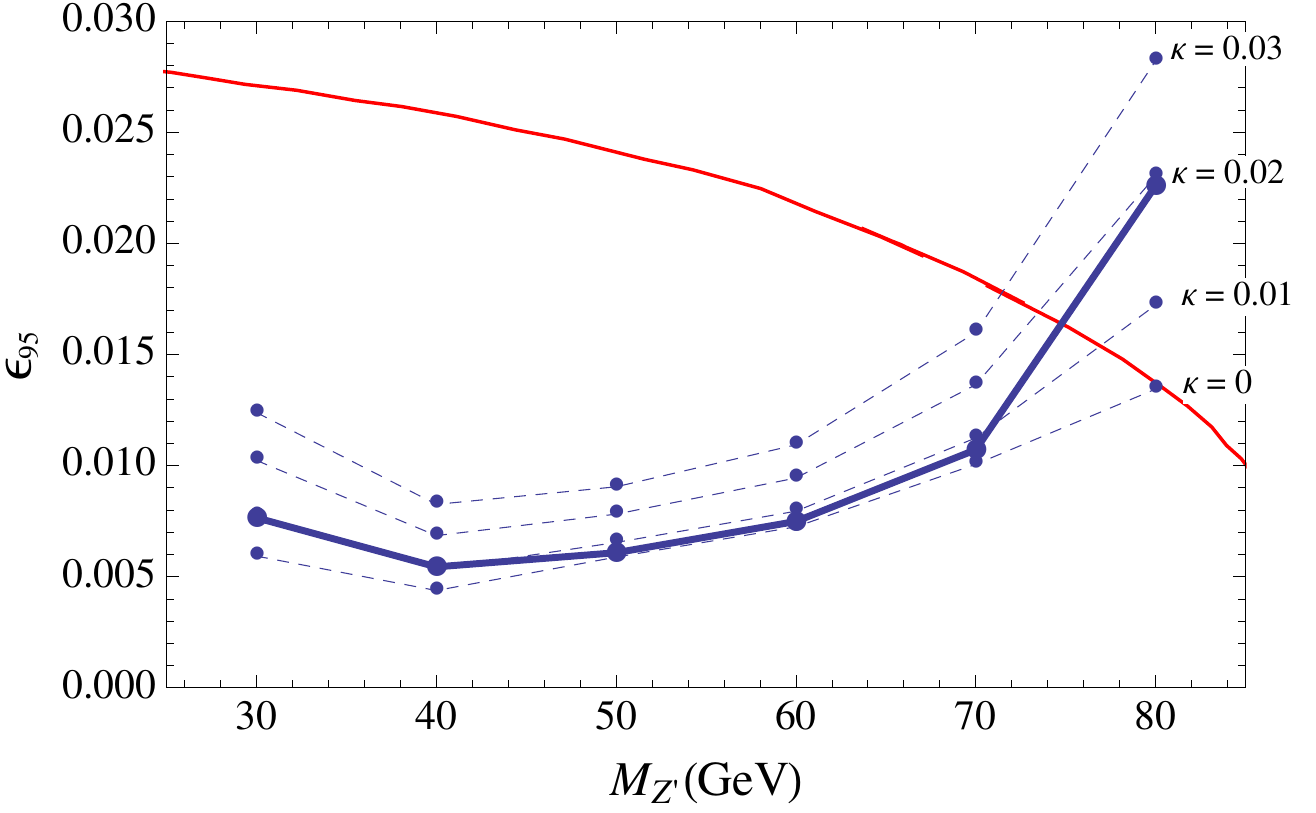}
\caption{Exclusion sensitivity estimates  for 7 TeV LHC data based on an integrated luminosity of 4.5 fb$^{-1}$, with the cut $(p_T)_{\mu \mu}>23$ GeV (left), and without a $(p_T)_{\mu \mu}$ cut (right).
For the thick blue line, $\kappa$ is set according to the procedure described in Section \ref{sec:background_model}.  The dashed lines show results for other $\kappa$ values.  The thin red line is the precision electroweak constraint taken from Ref.~\cite{Hook:2010tw}.
}
\label{fig:results_7TeV_1GeV_bins}
\end{center}
\end{figure}

 Our resulting sensitivity estimates are shown in Figure \ref{fig:results_7TeV_1GeV_bins}.  
For  the $\kappa$ values of Tables~\ref{tab:7TeV_14_9_0_fits} and \ref{tab:7TeV_14_9_23_fits}, the median values of $\epsilon_{95}$ go as low as a factor of four below the upper bound on $\epsilon$ from precision electroweak constraints.  

\section{Potential sensitivity of 8 TeV  and 14 TeV LHC data}
\label{sec:8TeV}
\begin{figure}[h]
\begin{center}
\includegraphics[width=4in]{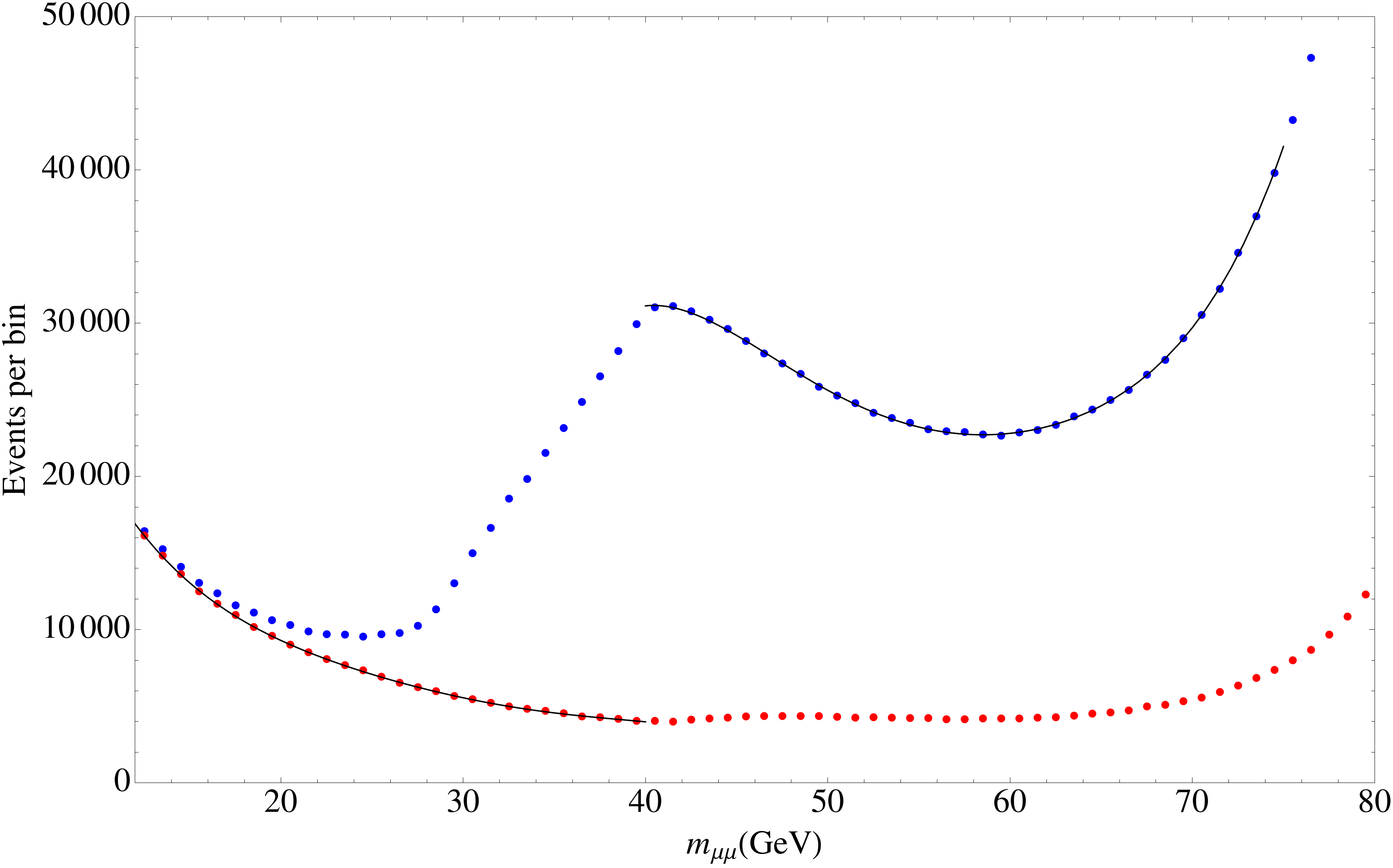}
\put(-130,110){\tiny no $\left( p_T \right)_{\mu \mu}$ cut}
\put(-250,28){\tiny $\left( p_T \right)_{\mu \mu}>30$ GeV}
\caption{For $\sqrt{s} = 8$ TeV and an integrated luminosity of 20 fb$^{-1}$, our simulated Drell-Yan $m_{\mu\mu}$ distributions with and without the $\left( p_T \right)_{\mu \mu}$ cut, along with best-fit fifth-order polynomial curves. The muons are required to have $\left(p_T \right)_1 > 20$ GeV, $\left(p_T \right)_2>10$ GeV, and $|\eta|<2.4$.}
\label{fig:8TeV_polynomial_fit_plot}
\end{center}
\end{figure}

At ${\sqrt s} = 8$ TeV,  the advantages of a larger integrated luminosity $\sim 20$ fb$^{-1}$ and slightly higher signal cross sections compete against higher muon $p_T$ thresholds required by the higher instantaneous luminosity.  Based on Ref.~\cite{Chatrchyan:2013iaa},  we conclude that it is realistic to consider hypothetical analyses based on muon $p_T$ cuts of 20 GeV and 10 GeV.  These cuts push the peak in the $m_{\mu \mu}$ invariant mass distribution to larger values of $m_{\mu \mu}$, as shown in Figure~\ref{fig:8TeV_polynomial_fit_plot}. 

As for the 7 TeV case, we consider two hypothetical analyses with 1-GeV $m_{\mu \mu}$ bins, one with  a $\left( p_T \right)_{\mu \mu}$ cut and one without.   We take the integrated luminosity to be $20$ fb$^{-1}$.   Fifth-order polynomial fits to our Drell-Yan Monte Carlo $m_{\mu \mu}$ distributions for the $12-40$ and  $40-75$ GeV mass ranges are shown in Figure~\ref{fig:8TeV_polynomial_fit_plot}.
 Tables~\ref{tab:8TeV_20_10_0_fits} and \ref{tab:8TeV_20_10_30_fits} show the $Z'$ masses tested for these analyses, the associated $m_{\mu \mu}$ ranges, and the $\kappa$ values determined the procedure described in Section \ref{sec:background_model}.

As before we apply a $75\%$ efficiency for the signal.  We normalize our Drell-Yan Monte Carlo simulations 
by requiring that the number of events in the $60-120$ GeV $m_{\mu \mu}$ mass range is equal to the number of events in that mass range for the 7 TeV case, multiplied by by three factors:  the ratio of the 8 and 7 TeV integrated luminosities,
 the ratio of the 8 and 7 TeV acceptances in that $m_{\mu \mu}$ range for the relevant muon $p_T$ and $\eta$ cuts, as determined by our simulations, and the ratio of the 8 and 7 TeV NNLO dimuon production cross sections in that $m_{\mu \mu}$ range.  We use $\sigma_\text{8 TeV} =  1.12$ nb and $\sigma_\text{ 7 TeV} =0.97$ nb, consistent with the values shown in Refs.~\cite{Chatrchyan:2014mua} based on calculations using FEWZ~\cite{Gavin:2010az}.
 
\begin{figure}[h]
\begin{center}
\includegraphics[width=3.in]{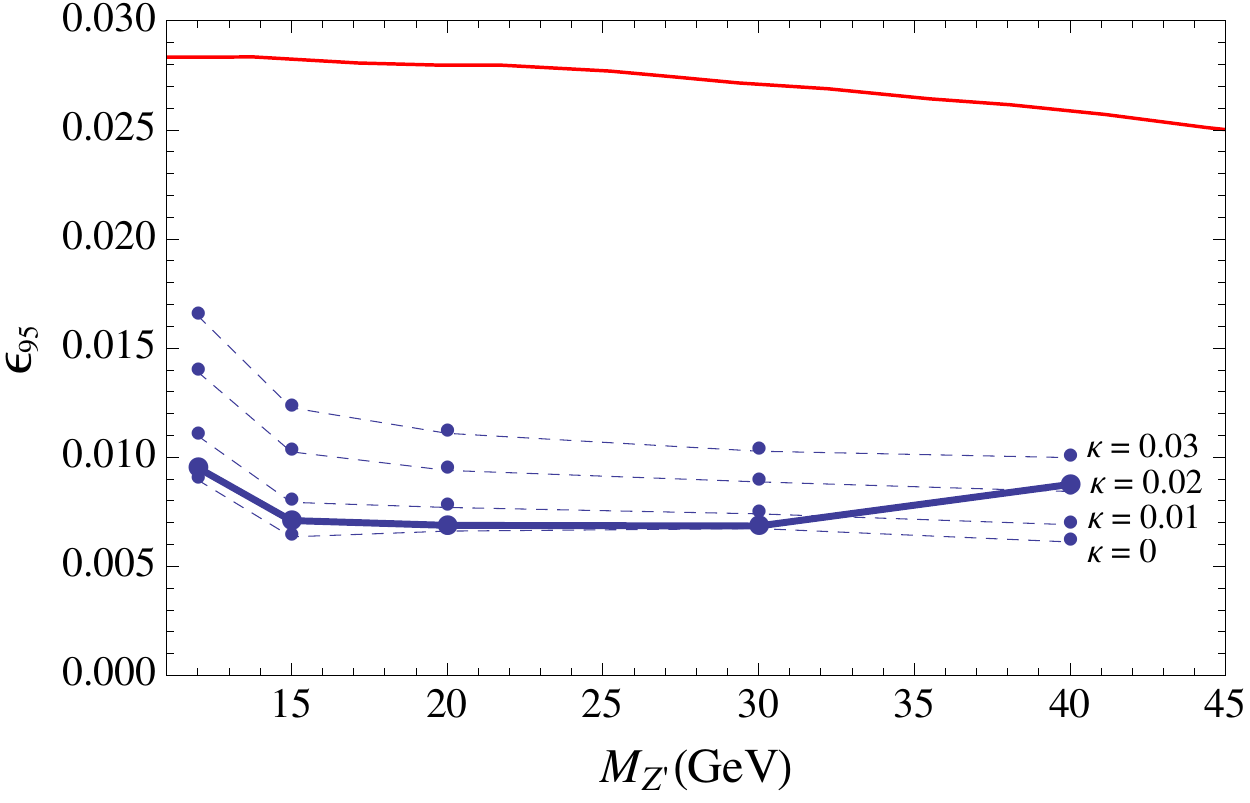}
\includegraphics[width=3.2in]{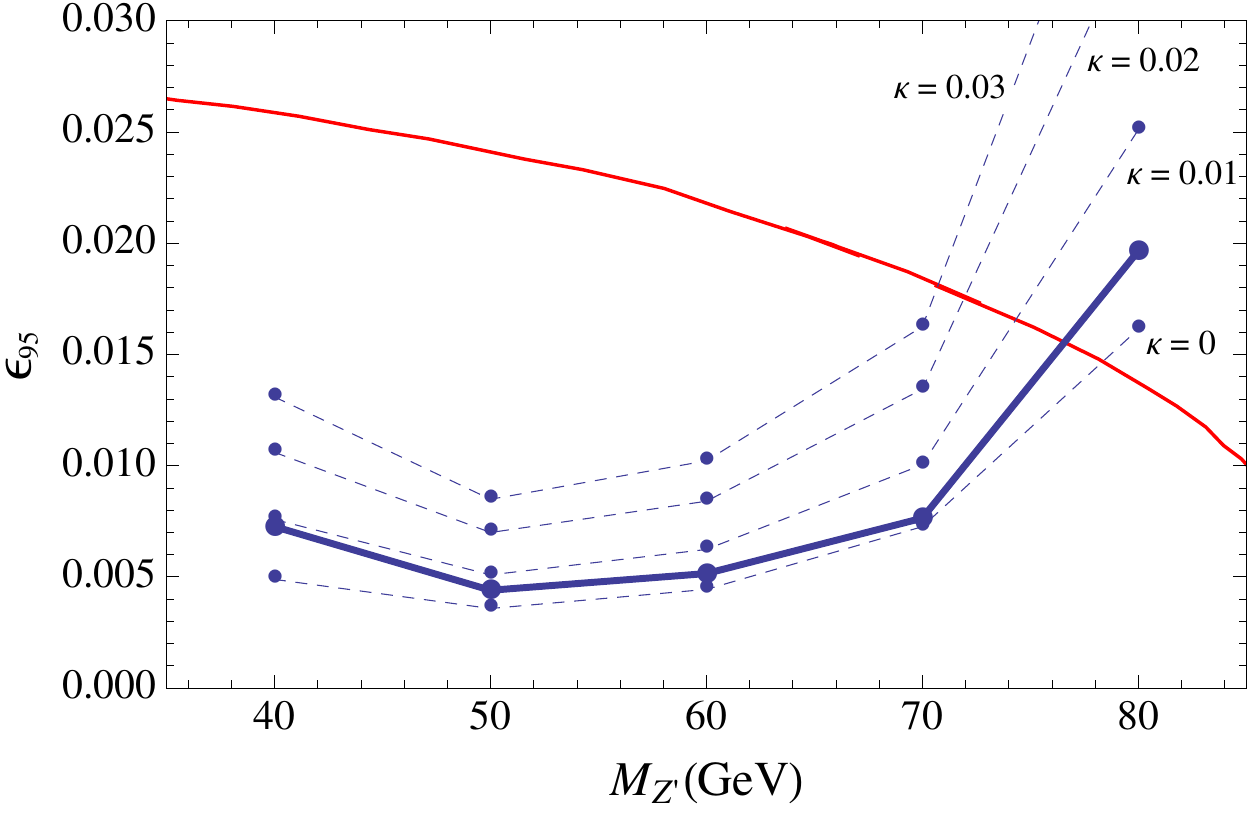}
\caption{Exclusion sensitivity estimates  for 8 TeV LHC data based on an integrated luminosity an integrated luminosity of 20 fb$^{-1}$, with the cut $(p_T)_{\mu \mu}>30$ GeV (left), and without a $(p_T)_{\mu \mu}$ cut (right).
For the thick blue line, $\kappa$ is set according to the procedure described in Section \ref{sec:background_model}.  The dashed lines show results for other $\kappa$ values.  The thin red line is the precision electroweak constraint taken from Ref.~\cite{Hook:2010tw}.
}
\label{fig:results_8TeV}
\end{center}
\end{figure}

Our sensitivity estimates are shown in Figure~\ref{fig:results_8TeV}.  These results suggest that, relative to the 7 TeV case, an analysis based on 8 TeV data would yield improved sensitivity for at least some of the $m_{\mu \mu}$ range below the $Z$ mass.  
\begin{figure}[h]
\begin{center}
\includegraphics[width=5.in]{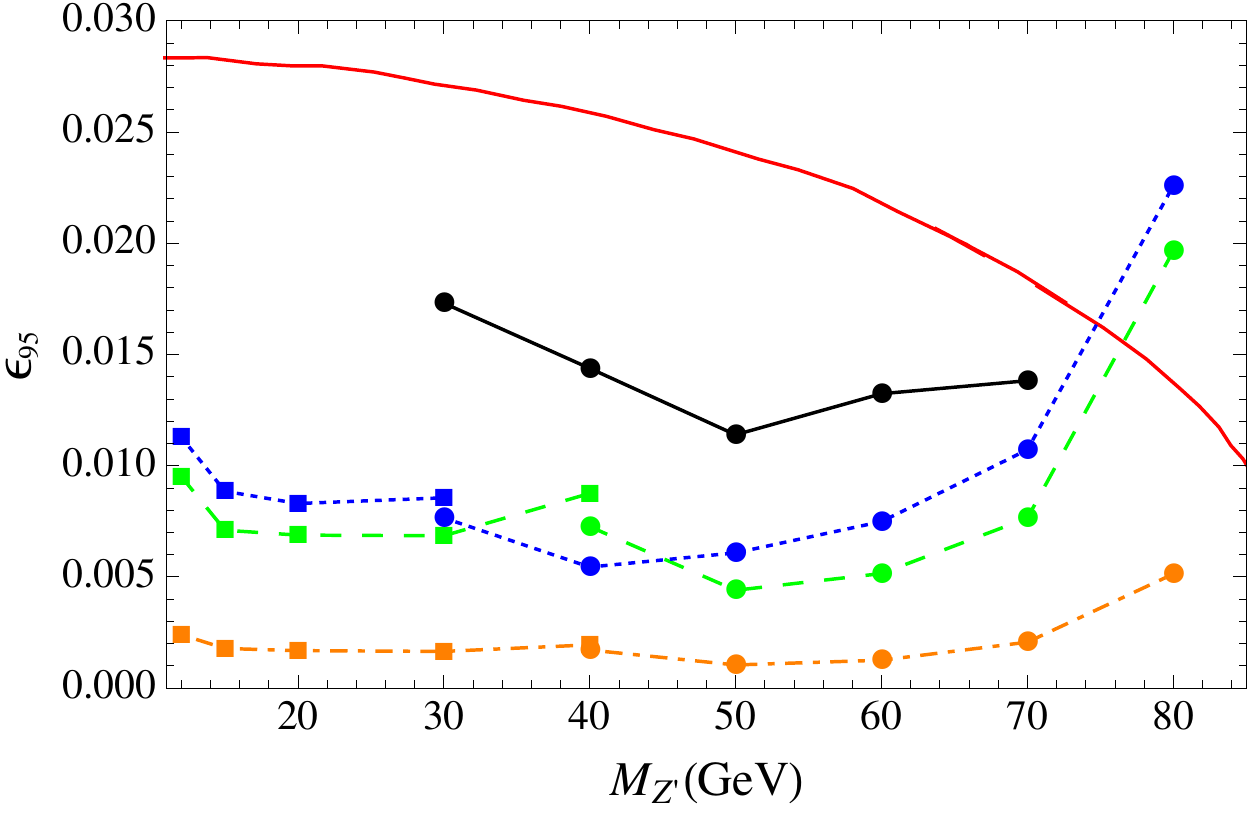}
\caption{Estimated upper bounds on $\epsilon$ based on the CMS analysis of Ref.~\cite{Chatrchyan:2013tia} (black, solid), and
 exclusion sensitivity estimates  for 7 TeV LHC data (blue, short dashed),  
 for 8 TeV LHC data (green, long dashed), 
 and, under the assumptions stated in the text, for 3000 fb$^{-1}$ of 14 TeV data (orange, dot-dashed). 
 Results for the low-mass analyses incorporating a $(p_T)_{\mu \mu}$ cut are indicated by square markers.  
 The $\kappa$ parameters are set according to the procedure described in Section \ref{sec:background_model}.  
 The thin red line is the precision electroweak constraint taken from Ref.~\cite{Hook:2010tw}.
 }
\label{fig:combined_results}
\end{center}
\end{figure}
Figure~\ref{fig:combined_results}
collects our results for estimated bounds on $\epsilon$ based on the 7 TeV CMS analysis along with estimates of potential sensitivity  for 7 and 8 TeV.

The ability of the 14 TeV LHC to probe relatively light dilepton resonances depends on trigger issues.    Refs.~\cite{ATLASHL,Tapper:2013yva} give reason to hope that various upgrades should allow trigger thresholds to be close to their 8 TeV values,  even at the High Luminosity LHC.   We will estimate the 14 TeV reach for the kinetically mixed $Z'$ model under the assumption that  the dimuon $p_T$ cuts we considered for 8 TeV analyses would be feasible at 14 TeV. 

For the mass range considered here, our simulations indicate that the the $\epsilon$ reach for 14 TeV and an integrated luminosity of 10 fb$^{-1}$ is approximately the same as the reach for 8 TeV and an integrated luminosity of 20 fb$^{-1}$.  
If we optimistically assume that the sensitivity for higher luminosities depends on $S/\sqrt{B}$, and therefore that the $\epsilon$ reach is proportional to $\left( \int \!{\mathcal L}\; dt\right)^{-1/4}$, we obtain the 14 TeV sensitivity estimates shown in  Figure~\ref{fig:combined_results} for an integrated luminosity of 3000 fb$^{-1}$.  These reach down to $\epsilon \sim 10^{-3}$.


\section{Conclusions}

LHC measurements of dilepton invariant mass distributions below $M_Z$ can be used as sensitives probe of new physics.  For the kinetically mixed $Z'$ model, analyses of 7 and 8 TeV data would be sensitive to $\epsilon$ values several times smaller than the upper bounds from precision electroweak constraints.  If trigger thresholds can be held near their 8 TeV values, $\epsilon$ values as small as $\sim10^{-3}$ may be accessible to the High Luminosity LHC.



\begin{acknowledgments}
 We are grateful to Kyle Cranmer, David Curtin, Stefania Gori, and Abi Sofer for helpful conversations.
This work was supported by NSF Grant \#1216168.    DTS thanks the Aspen Center for Physics and NSF Grant \#1066293 for hospitality as this paper was being completed.
\end{acknowledgments}

\newpage
\begin{table}[h]
   \centering
   \begin{tabular}{||c||c|c|c||} 
\hline \hline
$M_{Z'}$ (GeV)    & $\sigma_\text{7 TeV}$  (pb) & $\sigma_\text{8 TeV}$ (pb) & $\sigma_\text{14 TeV}$ (pb)\\
\hline
12       &  274   &  305 &  476 \\
15       &   163 &  182  &   288\\
20       &   80.6 &  90.3  &  147\\
30       &   28.5 &  32.4  &  54.6\\
40       &   13.3  &  15.3  &  26.3\\
50       &    7.43 &  8.56  & 15.1 \\
60       &    4.76 &  5.51  &  9.96\\
70       &    3.60 &  4.21  &  7.70\\
80       &     4.24&  4.94   &9.28                                    
 \\\hline \hline
   \end{tabular}
   \caption{Cross sections  for $pp \rightarrow Z' \rightarrow \mu^+ \mu^- $ used for our analyses, taking $\epsilon = 2\times 10^{-2}$.  The cross sections are proportional to $\epsilon^2$. }
   \label{tab:cross_sections}
\end{table}

\begin{table}[h]
   \centering
   \begin{tabular}{||c||c|c|c||} 
\hline \hline
$M_{Z'}$ (GeV)    & $m_{\mu \mu}$ range (GeV)&{\small median $p$-value, $\kappa \rightarrow 0$} & $\kappa$ \\
\hline
        30        & $30-76$    &  0.20 & $6.1\times 10^{-3}$\\
        40        & $35-76$    &  0.25 & $7.5\times 10^{-3}$\\
        50        & $35-76$    &  0.25 & $4.8\times 10^{-3}$\\
        60        & $35-76$    &  0.25 & $4.8\times 10^{-3}$\\
        70        & $35-76$    &  0.25 & $4.8\times 10^{-3}$\\
\hline \hline
   \end{tabular}
   \caption{Masses tested for the analysis based on the CMS results of Ref.~\cite{Chatrchyan:2013tia}, and associated fit information.  The third column gives the median $p$-value for the polynomial ($\kappa \rightarrow 0$) fit to pseudo data sets generated by statistically fluctuating the CMS Monte-Carlo-predicted SM bin counts reported in Figure 1 of Ref.~\cite{Chatrchyan:2013tia}.
   No $\left( p_T \right)_{\mu \mu}$ cut is imposed in this analysis.
   }
   \label{tab:7TeV_CMS_fits}
\end{table}
\begin{table}[h]
   \centering
   \begin{tabular}{||c||c|c|c||} 
\hline \hline
$M_{Z'}$ (GeV)    & $m_{\mu \mu}$ range (GeV)&{\small median $p$-value, $\kappa \rightarrow 0$} & $\kappa$ \\
\hline
        30        & $30-60$    &  0.26 & $9.6\times 10^{-3}$\\
        40        & $35-75$    &  0.27 & $1.1\times 10^{-2}$\\
        50        & $35-75$    &  0.27 & $4.9\times 10^{-3}$\\
        60        & $35-75$    &  0.27 & $4.9\times 10^{-3}$\\
        70        & $35-75$    &  0.27 & $7.2\times 10^{-3}$\\
        80        & $60-82$    &  0.21 & $1.9\times 10^{-2}$\\
\hline \hline
   \end{tabular}
   \caption{Masses tested for the $\sqrt{s} = 7$ TeV analysis based on 1-GeV bins and no 
$\left( p_T \right)_{\mu \mu}$ cut,  and associated fit information.  The third column gives the median $p$-value for the polynomial ($\kappa \rightarrow 0$) fit to pseudo data sets generated by statistically fluctuating our simulated Drell-Yan $m_{\mu \mu}$ distribution.}
   \label{tab:7TeV_14_9_0_fits}
\end{table}
\begin{table}[h]
   \centering
   \begin{tabular}{||c||c|c|c||} 
\hline \hline
$M_{Z'}$ (GeV)    & $m_{\mu \mu}$ range (GeV)&{\small median $p$-value, $\kappa \rightarrow 0$} & $\kappa$ \\
\hline
        12       & $12-40$    &  0.38 & $6.0\times 10^{-3}$\\
        15       & $12-40$    &  0.38 & $6.0\times 10^{-3}$\\
              20       & $12-40$    &  0.38 & $6.0\times 10^{-3}$\\
                      30       & $12-40$    &  0.38 & $9.2\times 10^{-3}$\\
\hline \hline
   \end{tabular}
   \caption{Masses tested for the $\sqrt{s} = 7$ TeV analysis based on 1-GeV bins with a
$\left( p_T \right)_{\mu \mu}>23$ GeV cut, and associated fit information.  The third column gives the median $p$-value for the polynomial ($\kappa \rightarrow 0$) fit to pseudo data sets generated by statistically fluctuating our simulated Drell-Yan $m_{\mu \mu}$ distribution.}
   \label{tab:7TeV_14_9_23_fits}
\end{table}

\begin{table}[h]
   \centering
   \begin{tabular}{||c||c|c|c||} 
\hline \hline
$M_{Z'}$ (GeV)    & $m_{\mu \mu}$ range (GeV)&{\small median $p$-value, $\kappa \rightarrow 0$} & $\kappa$ \\
\hline
        40        & $40-75$    &  0.28 & $9.1\times 10^{-3}$\\
        50        & $40-75$    &  0.28 & $6.6\times 10^{-3}$\\
        60        & $40-75$    &  0.28 & $5.3\times 10^{-3}$\\
        70        & $40-75$    &  0.28 & $2.6\times 10^{-3}$\\
        80        & $65-82$    &  0.38 & $5.0\times 10^{-3}$\\
\hline \hline
   \end{tabular}
   \caption{Masses tested for the $\sqrt{s} = 8$ TeV analysis with no 
$\left( p_T \right)_{\mu \mu}$ cut,  and associated fit information.  The third column gives the median $p$-value for the polynomial ($\kappa \rightarrow 0$) fit to pseudo data sets generated by statistically fluctuating our simulated Drell-Yan $m_{\mu \mu}$ distribution.}
   \label{tab:8TeV_20_10_0_fits}
\end{table}
\begin{table}[h]
   \centering
   \begin{tabular}{||c||c|c|c||} 
\hline \hline
$M_{Z'}$ (GeV)    & $m_{\mu \mu}$ range (GeV)&{\small median $p$-value, $\kappa \rightarrow 0$} & $\kappa$ \\
\hline
        12       & $12-40$    &  0.32 & $4.5\times 10^{-3}$\\
        15       & $12-40$    &  0.32 & $6.2\times 10^{-3}$\\
              20       & $12-40$    &  0.32 & $4.5\times 10^{-3}$\\
                      30       & $12-40$    &  0.32 & $4.5\times 10^{-3}$\\
                      40       & $12-45$    &  0.21 & $2.2\times 10^{-2}$\\
\hline \hline
   \end{tabular}
   \caption{Masses tested for the $\sqrt{s} = 8$ TeV  analysis with a
$\left( p_T \right)_{\mu \mu}>$ 30 GeV cut, and associated fit information.  The third column gives the median $p$-value for the polynomial ($\kappa \rightarrow 0$) fit to pseudo data sets generated by statistically fluctuating our simulated Drell-Yan $m_{\mu \mu}$ distribution.}
   \label{tab:8TeV_20_10_30_fits}
\end{table}

\end{document}